# Modified Opportunistic Deficit Round Robin Scheduling for improved QOS in IEEE 802.16 WBA networks

C.Kalyana Chakravarthy
Dept. of CSE
M.V.G.R.College of Engineering
Vizianagaram, India
kch.chilukuri@gmail.com

Dr. P.V.G.D.Prasad Reddy
Dept. of CS&SE
Andhra University College of Engineering
Visakhapatnam, India
prof.prasadreddy@gmail.com

*Abstract*— Packet and flow scheduling algorithms for WiMAX has been a topic of interest for a long time since the very inception of WiMAX networks. WiMAX offers advantages particularly in terms of Quality of service it offers over a longer range at the MAC level. In our work, we propose two credit based scheduling schemes one in which completed flows distributes the left over credits equally to all higher priority uncompleted flows(ODRREDC) and another in which completed flows give away all the excess credits to the highest priority uncompleted flow(ODRRSDC). Both the schemes are compatible with 802.16 MAC protocol and can efficiently serve real time bursty traffic with reduced latency and hence improved QOS for real time flows. We compare the two proposed schemes for their latency, bandwidth utilization and throughput for real time burst flows with the opportunity based Deficit Round Robin scheduling scheme. While the ODRR scheduler focuses on reducing the credits for the flows with errors, our approach also distributes these remaining credits together with the credits from completed flows equally among the higher priority uncompleted flows or totally to the highest priority uncompleted flow.

I. INTRODUCTION

IEEE 802.16 in PMP mode, defines five types of scheduling services[1] to support quality of service. They can be classified as Unsolicited Grant Services(UGS), Real-time Polling services(rtPS), Extended rtPS, non Real-time polling services(nrtPS) and Best Effort(BE).

Application of Unsolicited grant services (UGS) is Voice over IP (VoIP) without silence suppression. The mandatory service flow parameters that define this service are maximum sustained traffic rate, maximum latency, tolerated jitter, and request/transmission policy.

Applications of Real-time Polling service (rtPS) are Streaming audio and video, MPEG (Motion Picture Experts Group) encoded. The mandatory service flow parameters that define this service are minimum reserved traffic rate, maximum sustained traffic rate, maximum latency, and request/transmission policy.

Application of Extended real-time is VoIP with silence suppression. The mandatory service flow parameters are guaranteed data rate and delay.

Application of Non-real-time Polling service is File Transfer Protocol (FTP). The mandatory service flow parameters to define this service are minimum reserved traffic rate, maximum sustained traffic rate, traffic priority, and request/transmission policy.

Applications of Best-effort service (BE) are Web browsing, data transfer. The mandatory service flow parameters to define this service are maximum sustained traffic rate, traffic priority, and request/transmission policy.

In WiMAX, the MAC layer at the base station is fully responsible for allocating bandwidth to all users, in both the uplink and the downlink. The only time the MS has some control over bandwidth allocation is when it has multiple sessions or connections with the BS. In that case, the BS allocates bandwidth to the MS in the aggregate, and it is up to the MS to apportion it among the multiple connections. All other scheduling on the downlink *and* uplink is done by the BS. For the downlink, the BS can allocate bandwidth to each MS, based on the needs of the incoming traffic, without involving the MS. For the uplink, allocations have to be based on requests from the MS.

Different connection management strategies have been proposed, but the most common one is of management connections first, real-time connections followed by non-real time connections and finally Best Effort connections.

In our work, we propose and compare two credit based scheduling schemes, Opportunistic Deficit Round Robin Scheduling with Equal Distribution of Credits and Opportunistic Deficit Round Robin Scheduling with Single Distribution of Credits. The first one based on distribution of excess credits equally between all higher priority flows while the other proposed scheme is based on distribution of excess credits to the highest priority flow which is yet to be completed. The schemes are used to schedule flows between two classes of flows, real-time and non real-time flows.

We compare the two schemes in terms of the QOS parameters namely the throughput, bandwidth utilization, maximum latency etc., and observe that though the former one is based on fair scheduling, the latter in fact offers better performance under similar conditions compared to the opportunity-based DRR scheduling scheme.

## II. PREVIOUS WORK

A significant amount of work has already gone into scheduling disciplines that provide delay guarantees and fairness.

Time stamp scheduler essentially uses the idea of assigning time stamps to packets and then transmitting the packets in some order that achieves fairness. WFQ[3] and $WF^2Q$[4] algorithms fall into this category. However, both of the schemes require a reference with the GPS server to be maintained. Variants of WFQ include Self-Clocked FairScheduling [5] and Virtual Clock [6], which do not need to maintain a reference GPS server and hence can compute the time stamp in a more efficient way. Though time stamp schedulers have good delay properties, their processing time is quite high.

Round-robin schedulers [7][8][9][10] are the other broad class of work-conserving schedulers. These schedulers typically assign time slots to flows in some sort of round-robin fashion. Though they have better complexity compared to packet schedulers, however they have poor delay characteristics, particularly for packets of varying sizes.

Several improvements have been proposed to improve the delay properties of the basic Round-robin scheduler. There is another class of algorithms that try to combine the tight delay bound of time stamp based schedulers and the low time complexity of round robin based schedulers. They usually adopt a basic round robin like scheduling policy plus time stamp based scheduling on a reduced number of units [11]. Bin Sort Fair Queueing [12] is based on arranging packets into different bins based on their time stamps and scheduling in a FIFO manner.

Stratified Round Robin [13] uses the round robin approach for inter-class scheduling and the time stamp approach for intra-class scheduling after grouping flows into respective classes.

Recently proposed algorithms like ADRR [14] enhance the deficit round robin scheduling discipline by taking into account the channel quality experienced by the transmitting node. The ADRR scheduler is designed to achieve performance isolation among links characterized by heterogeneous channel conditions.

In the DRR scheme, Stochaic fair queuing is used to assign flows to queues. For servicing the queues, Round-robin servicing is used, with a quantum of service attached to each queue. It differs from the traditional Round-robin in that if a queue is unable to send a packet in the previous round because a packet was too large, the remainder from the previous quantum is added to the quantum for the next round. Queues that are not completely serviced in a round are compensated in the next round. However, once a flow is serviced, irrespective of its weight, it must wait for $N-1$ other flows to be serviced until it is serviced again. Also, during each round, a flow transmits its entire quantum at once. As a result, DRR has poor delay and burstiness properties.

The Smoothed Round Robin discipline addresses the output burstiness problem of DRR. This is done by spreading the quantum allocated to a flow over an entire round using a Weight Spread Sequence. Although SRR also results in better delay bounds than DRR, the worst case delay experienced by a packet is still proportional to $N$, the number of flows.

The Opportunity-based Deficit Round Robin scheduling scheme [16] is an improvement over the DRR scheme in that it considers the channel status in decisions it makes in serving flows. Opportunity-based scheduling ensures that a flow is not allowed to win a larger allocation of the resource if it uses its allocated resource inefficiently. ODRR used a penalty factor defined as *PenaltyFactor = $S_I / S_A$* to achieve fairness in allocation of resource S, where $S_I$ is the ideal number of bytes transmitted per unit of resource consumed and $S_A$ is the actual number of bytes transmitted per unit of resource S consumed.

With the use of the penalty factor, during an execution of the ODRR scheduler, a flow that takes longer to transmit a packet will have its deficit counter decremented by a larger amount than another flow that takes less time to successfully transmit a packet. As in DRR, when the packet to be transmitted next is predicted, the deficit counter is decremented and if it falls below 0, the scheduler proceeds to serve the next flow after adding the required quantum once it aborts transmission attempts from the current flow. The ODRR scheduler removes a packet from the queue only if its transmission succeeds.

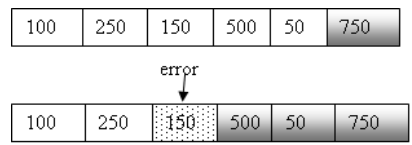

Fig. 1. The ODRR Scheduler

To understand how the ODRR scheduler works, consider the above scenario in Fig. 1. Since quantum size is 750, the packet with size 750 is processed in the first round. In the beginning of second round the credit count is incremented by remaining credits 0 plus quantum size 750.So in the second round, actually packets with sizes 50,500 and 150 can be served. However because of error in packet four, only 550 bytes can be served in the second round. Thus, the penalty factor = 550/700 =0.78571.

Thus the DC becomes DC=750-.78571*700= 750-550 = 200.Flow 2 is added to the queue. So when the flow2 is scheduled next, after other flows without errors are complete(not shown in figure), it receives only 200 + 750 credits, which is enough to send the packet number 4,together with the other packets successfully.

Clearly, the ODRR scheme is an improvement over the DRR scheme in terms of fairness and the throughput achieved. However, the problem with ODRR is that if a flow becomes backlogged due to errors, it has to wait until all the other subsequent flows without errors are completed before it is scheduled again. This may result in a large delay, and is undesirable, particularly if the flow is a high priority one and has encountered an error in the beginning of a round. The situation is similar to DRR[2] if all flows encounter errors in packet transmissions at the beginning each round.

The above problem could be considerably alleviated if we allow for distribution of the unused credits. It has to be understood that the distribution of credits occurs in case if a flow is complete and it has balance credits.

Our scheme is different from ODRR in the following aspects:

*1)*    *It provides for sharing of credits among uncompleted flows, allowing them to complete faster.*

*2)*    *In addition to reducing the credits for a flow with error, we try to increase the credits for other uncompleted flows without errors enabling them to complete faster.*

*3)*    *It does not consider the penalty factor when distributing credits.This reduces the overhead of determining the fraction of the packets transmitted with errors to those transmitted without error in case of an error, the balance credits are simply the quantum size minus the sum of credits of successfully transmitted packets, excluding the current packet with error.*

Our proposed algorithm effectively reduces the latency between flows while at the same time providing an improved throughput and ensuring fairness.

III. OPPURTUNISTIC DEFICIT ROUND ROBIN SCHEDULING WITH EQUAL DISTRIBUTION OF CREDITS (*ODRREDC*)

In our proposed algorithm, when a flow encounters an error, it is suspended and added to a queue. All the other completed flows distribute their balance credits among the higher priority flows. Thus all the higher priority flows that are incomplete in a round receive some additional credits, in addition to the quantum size in their next round.

This allows the other flows without errors to complete faster, which in turn reduces the delay for processing the flows with errors in the queues in the order of their priority.

Our model uses Inter-class scheduling for servicing the flows. It assumes fixed scheduling intervals between flows associated with a particular flow class. For each class $F_k$, the length of a scheduling interval is always $2^k$ slots. If a scheduling interval for $F_k$ starts at slot *t*, the next scheduling interval for $F_k$ starts at slot $t + 2k$, and so on. A flow is backlogged if it has not received it's fair share of bandwidth, i.e it still requires to be serviced in the next rounds. Backlogged flows are considered to be active. After every pending flow is serviced in the current time slot, clock time is $t_c$ is incremented. Otherwise, $t_c$ is advanced to the earliest time when some flow class becomes pending again.

Also, in our model the bandwidth is shared equally between the flows.

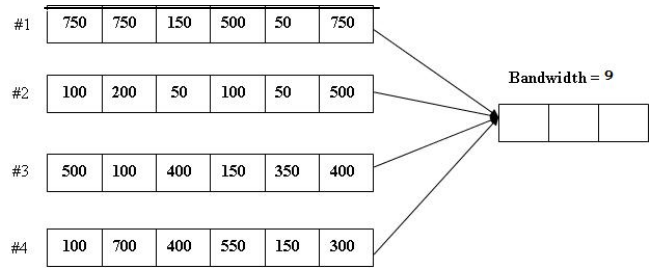

Fig. 2.  The Simulation Setup

The following scenario explains the operation of the ODRREDC. For simplicity, we have chosen the quantum size to be at least equal to the maximum packet size and the service pointer advances after each flow has been serviced.

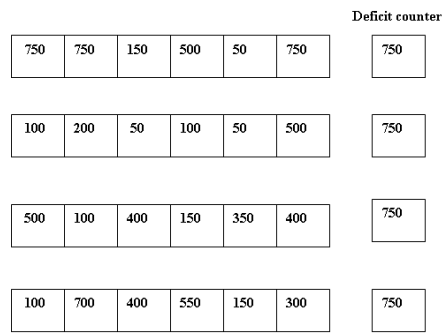

Fig. 3. Beginning of Round one

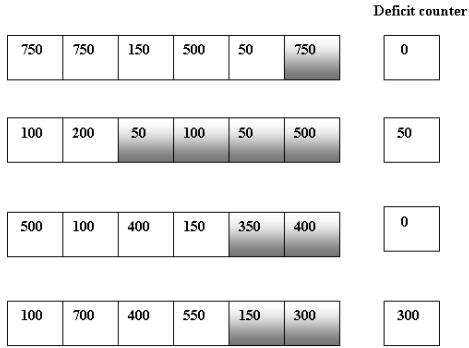

Fig. 4. End of Round one

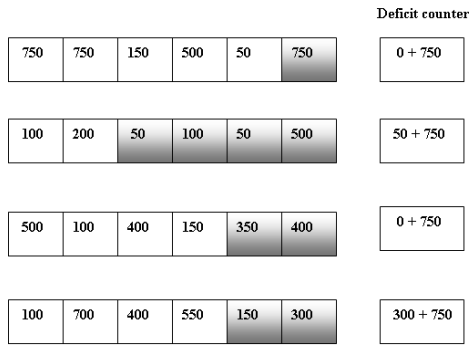

Fig. 5. Beginning of Round two

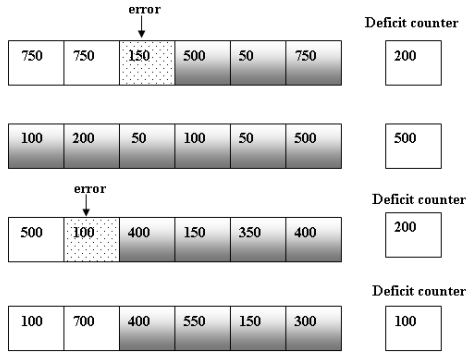

Fig. 6. End of Round two

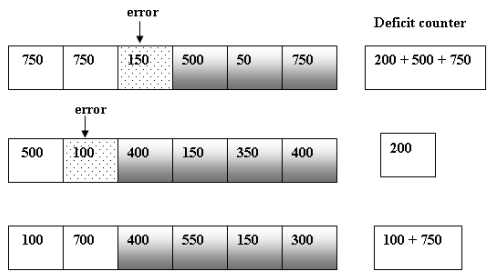

Fig. 7. Beginning of Round three

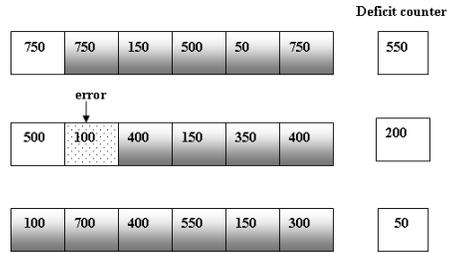

Fig. 8. End of Round three

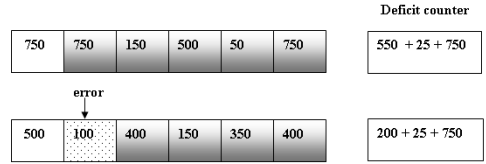

Fig. 9. Beginning of Round four for ODRREDC

At the beginning of round 4, the entire 650+750 credits go to the first flow in the queue with error, left over credits plus 750 go to second flow in the queue in the next(fifth) round. It can be observed that if flow 1 was not supplied with the excess credits, it would have taken one more additional round to complete.

It can be noted that flows which are complete donate their debit to the highest priority flows yet to be completed, while other flows proceed the same way as in the DRRscheme.

```
1. Initialize:
2  ActiveList = NULL;
3  Enqueue:
4  i = QueueInWhichPacketArrives;
5  if (ExistsInActiveList(i) == FALSE) then
6     Append queue i to ActiveList;
7     DCi = 0;
8  end if;
9  Dequeue:
10 while (TRUE) do
11    if (ActiveList ≠ NULL) then
12       Remove head of ActiveList, say flow i;
13       DCi = DCi + Qi;
14       while (QueueIsEmpty(i) == FALSE) do
15          p = HeadOfLinePacketInQueue(i);
16          if (Size(p) > DCi) then
17             break; /* escape from the inner while loop */
18          end if;
19          Transmit(p);
21          DCi = DCi – Size of packet successfully transmitted excluding
22          packet with error
23          if (TransmissionSucceeds(p) == TRUE) then distribute balance
24          credits equally between higher priority uncompleted flows
25          else
26          if (TransmissionSucceeds(p) == FALSE) then
27             break; /* escape from the inner while loop */
28          end if;
29          Remove packet p from queue i;
30       end while;
31       if (DCi < 0) then
32          DCi = 0;
33       end if;
34       if (QueueIsEmpty(i) == FALSE) then
35          Append queue i to ActiveList;
36       end if;
37    end if;
38 end while;
```

Fig. 10. The ODRREDC algorithm

## IV. OPPURTUNISTIC DEFICIT ROUND ROBIN SCHEDULING WITH SINGLE DISTRIBUTION OF CREDITS (*ODRRSDC*)

In our proposed algorithm, when a flow encounters an error, it is suspended, added to a queue. All the other completed flows distribute their balance credits to the highest priority pending flow. Thus highest priority flows that are incomplete in a round receive some additional credits, in addition to the quantum size in their next round.

This allows the other flows without errors to complete faster, which in turn reduces the delay for processing the flows with errors in the queues in the order of their priority.

The SCBSS differs from other scheduling schemes as in [15] where generally a completed flow distributes its credits continuously in subsequent rounds to the higher priority flows until it has no more credits to distribute.

ODRRSDC operates the same as ODRREDC till end of round three in Fig .8. However, at the beginning of round four, all 50 credits from the completed flow are given away to the highest priority flow 1. This is shown in the figure below.

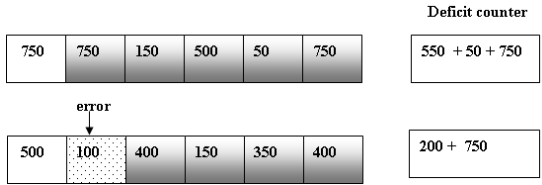

Fig. 11. Beginning of Round four for ODRRSDC

```
1. Initialize:
2  ActiveList = NULL;
3  Enqueue:
4  i = QueueInWhichPacketArrives;
5  if (ExistsInActiveList(i) == FALSE) then
6      Append queue i to ActiveList;
7      DCi = 0;
8  end if;
9  Dequeue:
10 while (TRUE) do
11     if (ActiveList ≠ NULL) then
12         Remove head of ActiveList, say flow i;
13         DCi = DCi + Qi;
14         while (QueueIsEmpty(i) == FALSE) do
15             p = HeadOfLinePacketInQueue(i);
16             if (Size(p) > DCi) then
17                 break; /* escape from the inner while loop */
18             end if;
19             Transmit(p);
20
21             DCi = DCi – Size of packet successfully transmitted excluding
22             packet with error
23             if (TransmissionSucceeds(p) == TRUE) then distribute balance
24             credits to the highest priority uncompleted flow
25             else
26             if (TransmissionSucceeds(p) == FALSE) then
27                 break; /* escape from the inner while loop */
28             end if;
29             Remove packet p from queue i;
30         end while;
31         if (DCi < 0) then
32             DCi = 0;
33         end if;
34         if (QueueIsEmpty(i) == FALSE) then
35             Append queue i to ActiveList;
36         end if;
37     end if;
38 end while;
```

Fig. 12. The ODRRSDC algorithm

## V. PROOFS

Theorem I: The proposed algorithm ODRREDC is better or atleast equal in terms of throughput and fairness compared to the ODRR.

Lemma 1: In an execution of the ODRR scheduler, at the end of round k,

$$0 \leq DC_i(k) \leq M \text{ for any flow } i, \text{ where } M \text{ is the maximum packet size.}$$

The lemma is identical to that in the case of DRR and the proof can be found in [2].

*Lemma 2:* During an execution of the ODRR scheduler over any m rounds, for any flow i,

$$mQ_i - M \leq SPT_i(M) \leq mQ_i + M$$

where $SPT_i(M)$ is the total potential throughput that can be achieved by a flow in m rounds, M is the maximum packet size and $Q_i$ is the quantum size to be added to a flow before the starting of each round. The potential throughput that a flow i may achieve during a round K is

$$PT_i(K) = Q_i + DC_i(k-1) - DC_i(k)$$

$$SPT_i(M) = \sum_{k=1}^{m} PT_i(K) = mQ_i + DC_i(0) - DC_i(m) \quad -1$$

Applying lemma 1, the statement of the lemma can be proved.

For our proposed method,

$$SPT_i^{'}(M) = \sum_{K=1}^{m} PT_i^{'}(K) = mQ_i^{'} + DC_i^{'}(0) - DC_i^{'}(m) \quad -2$$

Since in the proposed method, we are not reducing the quantum size given to packets with deficit credits,

$$Q_i^{'} \geq Q_i \text{ for all m} \quad -3$$

and since all completed flows distribute their balance credits to higher priority flows, for any round there exist flows with $DC_i^{'}(m) \leq DC_i(m)$. $\quad -4$

From equation 1-4, it can be deduced that
$SPT_i^{'}(M) \geq SPT_i(M)$ for some m

Theorem II: The proposed algorithm provides fairness atleast equal to that of the ODRR.

The fairness measure based on potential throughput measured across interval (t1,t2) is given by

$$FM(t_1,t_2) = \max_{V(i,j)} [SPT_i(t_1,t_2)/w_i - SPT_j(t_1,t_2)/w_j]$$

Where $w_i$ and $w_j$ are the weights of flows i and j respectively.

For our proposed method,
$$FM^{'}(t_1,t_2) = \max_{V(i,j)} [SPT_i^{'}(t_1,t_2)/w_i - SPT_j^{'}(t_1,t_2)/w_j]$$

It has already been proved that
$SPT_i(t_1,t_2)/w_i - SPT_j(t_1,t_2)/w_j \leq Q + 2M$ from lemma 2.
Since for our proposed method $Q = Q + \delta$, It follows that

$SPT_i(t_1,t_2)/w_i - SPT_j(t_1,t_2)/w_j \leq SPT_i^{'}(t_1,t_2)/w_i - SPT_j^{'}(t_1,t_2)/w_j$

⇨ FM(t1, t2) ≤ FM'(t1, t2) for any interval (t1,t2) which completes our proof.

## VI. SIMULATION RESULTS

We use a custom simulator written in java. The simulation runs in two threads - the flow generator that generates packet and the scheduler that checks at every configurable scheduling period and schedules the packets. Both these modules can be run either concurrently or independently. Simulation has been carried out on 20 queues, each containing maximum packets of variable size, for different quantum sizes for 20 seconds and the results have been evaluated. The packets are generated according to Poisson arrival process. For our results we limited the number of flows so that the sum total of their minimum bandwidth requirements matches the maximum capacity of the network. We have chosen a 9KbpS output queue and each input queue has six packets of maximum size 750 bytes and has a bandwidth of 4500 bps. All flows are critical and are arranged in the decreasing order of their priorities.

Our algorithm has shown reasonable improvement in terms of latency of critical flows, which makes it suitable for real time communications such as real time Video-on demand. If all latency critical flows meet the requirements, the maximum delay between latency critical flows should not exceed (n * s) + Max/B where n is number of latency critical flows, B bandwidth of the output line, s is maximum size of packet in a flow, Max is maximum quantum size.

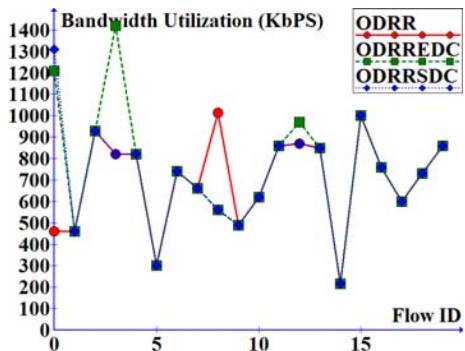
Fig. 13. Flow ID vs bandwidth utilization

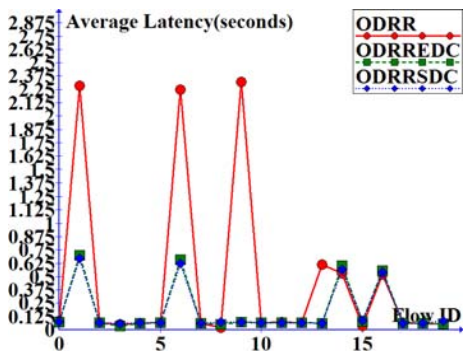
Fig 14. Flow ID vs Average latency

## VII. CONCLUSIONS

In our work, we have proposed two scheduling schemes ODRREDC and ODRRSDC for scheduling real time flows. It was observed from the results that while both the schemes perform better compared with the Opportunistic Deficit Round Robin scheduling scheme, the former is more suitable for real time flows under unsteady traffic conditions. In our method, any excessive idle bandwidth is reallocated to avoid wasting of available transmission capacity. In both cases, we assume scheduling under erroneous channel conditions.

Scheduling on Multiple Input Multiple Output channels with multiple antennas, scheduling on multi-hop networks for end to end service guarantees are also areas that demand further improvement.

### REFERENCES

[1] A. Sayenko, O. Alanen, J. Karhula and T. Hämäläinen," Ensuring the QoS Requirements in 802.16 Scheduling", Proceedings of IEEE/ACM MSWiM 2006, Torremolinos, Spain, Oct. 2006.

[2] M. Shreedhar and G.Varghese,"Efficient fair queuing using deficit Round Robin", in Proc. SIDCOMM '95 Boston, MA, Aug 1995.

[3] Demers,A.S Keshavand S.Shenkar,1989,"Analysis and simulation of a fair queuing problem", Proceedings of the Symposium and Communications Architectures and protocols, September 25-27,A.C.M NewYork, USA pp:1-12.

[4] Bennet , J.C.R and H.Zhang,1996,"WF2Q: Worst –case fair weighted fair queuing", Proceedings of the INFOCOM, March 24-28,San Fransisco,CA, pp:1-9

[5] Golestani,S.J.1994,"A self clocked fair queuing scheme for broad band applications", Proceedings of the 13[th] IEEE INFOCOM'94, Networking for Global Communications, June 12-16, Toronto, Ont, Canada, pp: 636-646.

[6] L. Zhang, "A new architecture for Packet switched network protocols", PhD dissertation, Massachesets Institute of technology, July 1989

[7] Lenzini, L., Mingozzi, E., and Stea, G. Aliquem: "a novel DRR implementation to achieve better latency and fairness at O(1) complexity," In IWQoS'02 (2002).

[8] "The Smoothed Round-Robin Scheduler", Paul Southerington, Member, IEEE, ECE742, 28 APRIL 2005.

[9] B. Bensaou, K. Chan, and D. Tsang, "Credit-based fair queuing (CBFQ): A simple and feasible scheduling algorithm for packet networks", IEEE ATM97 Workshop, pp. 589594, May 1997.

[10] Dessislava Nikolova and Chris Blondia, "Last-Backlogged First-Served Deficit Round Robin (LBFS-DRR) Packet Scheduling Algorithm", 15[th] IEEE International conference on networks,Nov.2007.

[11] Deng Pan, Yuanyuan Yang, "Credit Based Fair Scheduling for Packet Switched Networks", IEEE INFOCOM'05.

[12] S. Cheung and C. Pencea, "BSFQ: bin sort fair queuing," IEEE INFOCOM '02, pp. 1640-1649, New York, Jun. 2002.

[13] S. Ramabhadran, J. Pasquale, "Stratified round robin: a low complexity packet scheduler with bandwidth fairness and bounded delay," ACM SIGCOMM '03, pp. 239-250, Karlsruhe, Germany, Aug. 2003.

[14] Riggio, R. Miorandi, D. Chlamtac, "Airtime Deficit Round Robin (ADRR) packet scheduling algorithm," 5[th] IEEE International conference on Mobile Ad Hoc and Sensor Systems-MASS'08.

[15] Tsung-Yu Tsai, Zsehong Tsai, "Design of a Packet Scheduling Scheme for Downlink Channel in IEEE 802.16 BWA Systems", in WCNC'08 (2008).

[16] Yunkai Zhou, Madhusudan Hosaagrahara and Harish Sethu," Opportunity-Based Deficit Round Robin: A Novel Packet Scheduling Strategy for Wireless Networks", in Proceedings of the IEEE Workshop on High Performance Switching and Routing Kobe, Japan, May 26–29, 2002

AUTHORS PROFILE

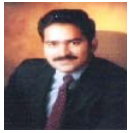

C.Kalyana Chakravarthy has a teaching experience of over nine Years and is currently working as Associate prof. in M.V.G.R. College of Engineering, Vizianagaram. He has been actively working on diverse areas of network caching MANETs routing protocols, resource allocation and scheduling in WiMAX, Mesh networks.

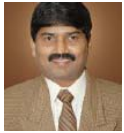

Dr. P.V.G.D Prasad Reddy has a teaching experience of over twenty years. He is currently serving as the Registrar at the Andhra University. He has over 16 publications in International Journals and 20 papers in conferences. His Research areas include Soft Computing, Software Architectures, Knowledge Discovery from Databases , Image Processing , Number theory & Cryptosystems.